\documentstyle[amssymb,aps,prl,twocolumn,epsfig]{revtex}

\begin{document}

\twocolumn[\hsize\textwidth\columnwidth\hsize\csname
@twocolumnfalse\endcsname

\title{Hall effect in c-axis-oriented MgB$_{2}$ thin films}
\author{W. N. Kang,$^*$ Hyeong-Jin Kim, Eun-Mi Choi, Heon
Jung Kim, Kijoon H. P. Kim, H. S. Lee, and Sung-Ik Lee}

\address{National Creative Research Initiative Center for Superconductivity
and Department of Physics, Pohang University of Science and
Technology, Pohang 790-784, Republic of Korea } \draft \maketitle

\begin{abstract}
We have measured the longitudinal resistivity and the Hall
resistivity in the ab-plane of highly c-axis-oriented MgB$_{2}$
thin films. In the normal state, the Hall coefficient ($R_{H}$)
behaves as $R_{H}\thicksim T$ \ with increasing temperature ($T$)
up to 130 K and then deviates from that linear $T$-dependence at
higher temperatures. The $T^{2}$ dependence of the
cotangent of the Hall angle is only observed above 130 K. The
mixed-state Hall effect reveals no sign anomaly over a wide range
of current densities from 10$^{2}$ to 10$^{4}$ A/cm$^{2}$ and for
magnetic fields up to 5 T.
\end{abstract}

\vskip 1.5pc]

\section{INTRODUCTION}

The recent discovery of superconductivity in MgB$_{2}$, with a transition
temperature ($T_{c}$) of about 39 K\cite{Nagamatsu01}, has attracted
extensive scientific interest in the fields of basic research and
applications. Already, several physical properties, such as the Hall effect,
the thermoelectric power, the magnetization, and the magnetoresistance, have
been investigated using polycrystalline samples \cite
{Kang01,Lorentz01,Choi01,MSKim01,Finnemore01}. However, many experimental
results still remain controversial because of the relatively high
anisotropic nature of this compound. Based on measurements of the upper
critical field for different crystal planes of MgB$_{2}$ single crystals 
\cite{KJKim01,SLee01} and of highly c-axis-oriented thin films\cite{Jung01},
they have confirmed the anisotropic nature of MgB$_{2}$ superconductor.
These results strongly suggest that the physical properties of MgB$_{2}$
should be investigated by either using single crystals or high-quality thin
films having preferred orientations. For the Hall measurements, since
sizable single crystals are not available and the Hall signal is very small
due to its metallic character, the thin-film form is the best candidate for
achieving accurate experimental results.

In our earlier work on polycrystalline samples \cite{Kang01}, we confirmed
that the majority charge carriers were holelike, which was consistent with
theoretical estimates \cite{Satta01}; subsequently, similar results were
also reported for polycrystalline MgB$_{2}$ thin films \cite{Jin01}. To the
best of our knowledge, the in-plane Hall effect for MgB$_{2}$ has not been
previously studied; thus, measurement of the ab-plane Hall effect for
c-axis-oriented MgB$_{2}$ thin films should provide significant input for
future investigations of its electronic transport properties and vortex
dynamics.

For high-$T_{c}$ cuprate superconductors (HTS), a universal $T^{2}$
dependence of the cotangent of the Hall angle (cot$\Theta _{H}$) has been
extensively discussed. Anderson\cite{Anderson91} has proposed that this
behavior can be explained if two different scattering rates, a transport
scattering time and a Hall (transverse) scattering time, are considered,
where the longitudinal resistivity ($\rho _{xx}$) is determined by the
former and the Hall resistivity ($\rho _{xy}$) is determined by both. Most
experimental results for HTS have supported this theory \cite
{Abe99,Chien91,Carrington92}, and it is generally accepted that, in the
normal state, a cot$\Theta _{H}$ $\sim $ $T^{2}$ law is universal over a
wide temperature range. Similar behavior has also been reported for
polycrystalline MgB$_{2}$ superconductors\cite{Kang01,Jin01}.

Another interesting feature concerning the mixed-state Hall effect as a
probe of superconductivity is the anomalous sign change near $T_{c}$ as a
function of the $T$ and the magnetic field, and its origin has remained an
unsolved subject for over 30 years. The sign anomaly has been observed in
some conventional superconductors \cite{Hagen93}, as well as in most HTS 
\cite{Hagen93,Kang96,Kang00}. However, in clean superconductors, such as
pure Nb, V, and 2{\it H}-NbSe2, no sign anomaly has been found \cite{Hagen93}%
. Our Hall data for MgB$_{2}$ are more similar to the behavior seen in Nb,
V, and 2{\it H}-NbSe2, suggesting that MgB$_{2}$ might be a clean-limit
superconductor \cite{Canfield01}.

In this paper, we report the first measurement of the in-plane Hall effect
of MgB$_{2}$. The measurement was carried out using highly c-axis-oriented
thin films, and we found that the sign of the $R_{H}$ was positive like
those of HTS. Also, the $R_{H}$ appeared to follow a linear behavior for the 
$T$ region from 30 to 130 K, which is different from the behaviors of
polycrystalline MgB$_{2}$ and of HTS. The Hall effect in the mixed state
showed no sign anomaly over a wide range of current densities from 10$^{2}$
to 10$^{4}$ A/cm$^{2}$ and for magnetic field up to 5 T, which is contrast
to the observations in most HTS and polycrystalline MgB$_{2}$ thin films.

\bigskip

\section{EXPERIMENT}

The MgB$_{2}$ thin films were grown on Al$_{2}$O$_{3}$ (1 $\bar{1}$ 0 2)
single crystals under a high-vacuum condition of $\sim $ 10$^{-7}$ Torr by
using the pulsed laser deposition and the postannealing techniques reported
in an earlier paper \cite{Kang01b}. Typical samples were 10 mm in length, 10
mm in width, and 0.4 $\mu $m in thickness. The film thickness was measured
using scanning electron microscopy. Standard photolithographic techniques
were used to produce thin-film Hall bar patterns, which consisted of a
rectangular strip (1 mm in width and 3 mm in length) of MgB$_{2}$ with three
pairs of sidearms (the upper inset of Fig. 2). The narrow sidearm width of
0.1 mm was patterned so that the sidearms would have an insignificant effect
on the equipotential. Using this 6-probe configuration, we were able to
measure simultaneously the $\rho _{xx}$ and $\rho _{xy}$ at the same $T$;
thus the cot$\Theta _{H}$ was obtained very precisely. To achieve good ohmic
contacts (%
\mbox{$<$}%
1 $\Omega $), we coated Au film on the contact pads after cleaning the
sample surface by using Ar-ion milling. After installing a low-noise
preamplifier prior to the nanovoltmeter, we achieved a voltage resolution of
below 1 nV. The magnetic field was applied perpendicular to the sample
surface by using a superconducting magnet system, and the applied current
densities were 10$^{2}-$10$^{4}$ A/cm$^{2}$. The Hall voltage was found to
be linear in both the current and the magnetic field.

\bigskip

\section{RESULTS and DISCUSSION}

The structural analysis was carried using X-ray diffractometry, and the
results are shown in Fig. 1(a). The MgB$_{2}$ thin film showed a highly
c-axis-oriented crystal structure, and the sample purity exceeded 99$\%$ and
had only a minor \{101\} oriented phase. Figure 1(b) shows the low-field
magnetization at $H=4$ Oe for both the zero-field-cooled (ZFC) and
field-cooled (FC) states of an MgB$_{2}$ film. A very sharp diamagnetic
transition is observed. Even at a high $T$ of 37 K and under self-field
conditions, the critical current density determined by direct current vs
voltage measurements was observed to be on the order of 10$^{5}$ A/cm$^{2}$ 
\cite{HJKim01}. These results indicate that the MgB$_{2}$ films used in the
present study were homogeneous and of very high quality.

Figure 2 shows the $T$ dependence of $\rho _{xx}$ for a MgB$_{2}$ film at H
= 0 and 5 T. The upper inset shows the 6-probe Hall-bar pattern. Pads 3 - 4
and 5 - 6 were used to measure $\rho _{xy}$ and $\rho _{xx}$, respectively,
while the current was applied between 1 and 2. The lower inset shows a
magnified view near the superconducting transition. The onset $T_{c}$ was
39.2 K and had a narrow transition width of $\sim $ 0.15 K, as judged from
the 10 to 90$\%$ superconducting transition. At 40 K, $\rho _{xx}$ was 3.4 $%
\mu \Omega $ cm, giving a residual resistivity ratio [RRR = $\rho _{xx}(300$K%
$)/\rho _{xx}(40$K)] of 3, which was smaller than the value observed for MgB$%
_{2}$ single crystals\cite{KJKim01,SLee01}. A very small (less than 0.5$\%$)
magnetoresistance was observed in the normal state at 5 T.

The $T$ dependence of the $R_{H}$ at 5 T is shown in Fig. 3. The offset
voltage due to the misaligned Hall electrodes was eliminated by reversing
the field from -5 T to 5 T (inset of Fig. 3), and the Hall voltage was taken
as the average value, V$_{xy}$ = (V$_{+H}$ - V$_{-H}$)/2, for all data
points. The offset voltage at H = 0 T was very small compared to V$_{xy}$ at
5 T, indicating excellent alignment of the Hall electrodes. The value of V$%
_{xy}$ was about 2 orders of magnitude larger than that of polycrystalline
bulk samples \cite{Kang01}. Due to our high-resolution measurements, we were
able to interpret our Hall data rigorously . The value of $R_{H}$ was
positive over the entire $T$ range, which is consistent with the result of
band calculations\cite{Satta01}. Although the charge carrier density cannot
be determined simply within the context of the Boltzmann theory because of
the anisotropic band structure and the complex Fermi surface of MgB$_{2}$,
such a calculation would be meaningful for comparison with other
superconductors. At 100 K, the $R_{H}$ was 3 $\times $ 10$^{-11}$ m$^{3}$/C,
and the hole carrier density, calculated from $1/eR_{H}$, was $\thicksim $ 2 
$\times $ 10$^{23}$ holes/cm$^{3}$. The absolute value of the hole density
was two orders of magnitude larger than that YBa$_{2}$Cu$_{3}$O$_{7}$ \cite
{Harris92}, indicating that MgB$_{2}$ has a metallic superconductor. The
average value of the $R_{H}$ was consistent with theoretical estimates\cite
{Satta01}.

As the $T$ was increased from the $T_{c}$, the $R_{H}$ decreased linearly up
to 130 K (T$^{\ast }$) and then deviated from that linear behavior at higher 
$T$, suggesting that the electronic transport mechanism changes at around
130 K. The T$^{\ast }$ was observed to be independent of magnetic fields up
to 5 T. This feature is somewhat different from previous results for
polycrystalline bulk \cite{Kang01} and thin-film samples \cite{Jin01} for
which the $R_{H}$ exhibited the same $T$ dependence over the entire $T$
range from the $T_{c}$ to 300 K. These results suggest that MgB$_{2}$ might
have different transport mechanisms in the in-plane and the out-of-plane
directions. A similar distinct $T$-dependence around 150 K was also observed
in the thermoelectric power measurements \cite{Lorentz01,Choi01}; the
thermoelectric power increased linearly with increasing $T$ up to around 150
and then showed a downward deviation from linearity. This behavior is
believed to be due to the different $T$ dependences of the multi-band
contributions to the transport properties; at low $T$ (below 150 K), charge
transport is governed mainly by hole carriers whereas at higher $T$, the
contribution of electron carriers must be considered\cite{Lorentz01}. \ 

In Fig. 4, we show the $T$ dependence of cot$\theta _{H}$ at 5 T. A good
linear fit to $AT^{2}+B$ is observed for the $T$ range from 130 to 300 K;
clear deviation from a $T^{2}$ dependence is seen below 130 K. According to
the Anderson theory\cite{Anderson91}, which is based on charge-spin
separation, charge transport is governed by two separate scattering times
with different $T$ dependences. The longitudinal conductivity ($\sigma _{xx}$%
) is proportional to the transport scattering time ($\tau _{tr}$) whereas
the Hall conductivity ($\sigma _{xy}$) is determined by $\tau _{H}\tau _{tr}$
where the Hall relaxation time ($\tau _{H}$) is proportional to $1/T^{2}$.
The $\tau _{H}$ is mainly governed by spinon-spinon interactions; thus, its $%
T$ dependence is not affected by impurities. As a result, the cot$\theta
_{H} $ (= $\sigma _{xx}/\sigma _{xy}$) should follow a $T^{2\text{ }}$law.
Such a universal temperature dependence has been observed in most HTS\cite
{Abe99}, and the $T^{2\text{ }}$law has been confirmed not to depend on
impurities \cite{Chien91,Carrington92}. Above 130 K, our experimental data
are also in good agreement with a $T^{2\text{ }}$law as observed in most
HTS. However, our data cannot be interpreted within the Anderson theory
because MgB$_{2}$ does not have active spins. We also observe a distinct $%
T^{2}$ dependence of cot$\theta _{H}$ at around 130 K.

Finally, we address the transport properties in the superconducting state.
In Fig. 5(a), we show the $T$ dependence of the $\rho _{xx}$ for magnetic
fields of 2 and 5 T and at current densities of 10$^{2}$ to 10$^{4}$ A/cm$%
^{2}$. A broad superconducting transition is observed, which implies the
existence of a relatively wide vortex-liquid phase in this compound. This
result is quite similar to those for HTS. In a separate paper\cite{HJKim01b}%
, we reported that this vortex phase could be interpreted well by using two
distinct regions; a thermal fluctuation region at high $T$ near T$_{c}$ and
a vortex-glass region at low $T$. Moreover, we found a very narrow thermally
activated flux-flow region, which was different from the case of HTS.

The corresponding $\rho _{xy}$ data are plotted in Fig. 5(b). No sign
reversal was detected in the Hall data measured for magnetic fields from 1
to 5 T and over two orders of magnitude of the current density. A puzzling
sign reversal has been observed in the mixed-state Hall effect for most HTS 
\cite{Kang00} and even for polycrystalline MgB$_{2}$ films\cite{Jin01}. In
conventional superconductors, this sign change occurs mostly in moderately
clean superconductors, but is not seen in either clean-limit
superconductors, such as V, Nb, and NbSe$_{2}$, or dirty-limit
superconductors, such as superconducting alloys\cite{Hagen93}. Therefore,
the absence of the Hall sign anomaly suggests that MgB$_{2}$ should be
categorized as a clean-limit superconductor. Indeed, a short superconducting
coherence length ( $\sim $50 \r{A}) and a relatively large mean free path
(250 $-$ 600 \r{A}) have been reported for this compound \cite
{Canfield01,Manzano01}.

An interesting microscopic approach based on the time-dependent
Ginzburg-Landau theory has been proposed in a number of papers \cite
{Dorsey92,Kopnin93,Otterlo95}. According to this model, the mixed-state Hall
conductivity in type II superconductors is determined by the quasiparticle
contribution and the hydrodynamic contribution of the vortex cores. Since
the hydrodynamic contribution is determined by the energy derivative of the
density of states \cite{Kopnin93,Otterlo95}, if that term is negative and
dominates over the quasiparticle contribution, a sign anomaly can appear.
This theory is consistent with experimental data for HTS \cite{Ginsberg95}.
For the mixed-state Hall effect in MgB$_{2}$ compound, since no sign anomaly
was detected, we may suggest that the hydrodynamic contribution is very
small or negligible in this superconductor.

\bigskip

\section{SUMMARY}

Using high quality c-axis-oriented MgB$_{2}$ thin films, we studied the
in-plane Hall effect as a function of the magnetic field over a wide range
of current densities. The normal-state $R_{H}$ increased linearly with
increasing $T$ up to 130 K and then showed a downward deviation from its
linear dependence at higher $T$, which is probably due to the distinct $T$
dependences of the electronic states of the MgB$_{2}$ compound. Our Hall
data were also in good agreement with a cot$\theta _{H}\thicksim T^{2\text{ }%
}$law above 130 K. The mixed-state Hall effect revealed no sign anomaly for
magnetic fields from 1 to 5 T over two orders of magnitude of the current
density.

ACKNOWLEDGMENTS: We thank J. Crayhold for useful discussions. This work is
supported by the Ministry of Science and Technology of Korea through the
Creative Research Initiative Program.

\begin{figure}[tbp]
\caption{(a) X-ray diffraction patterns of MgB$_{2}$ thin films. A highly
c-axis-oriented crystal structure normal to the substrate surfaces was
observed. S denotes the substrate peaks. (b) Magnetization at $H=4$ Oe in
the ZFC and FC states.}
\label{fig:fig1}
\end{figure}

\begin{figure}[tbp]
\caption{Temperature dependence of the resistivity of MgB$_{2}$ thin films
for H = 0 and 5 T. The lower inset shows a magnified view near the $T_{c}$.
A sharp transition, with a narrow transition width of $\sim $ 0.15 K, was
observed. The upper inset is a schematic diagram of the Hall bar pattern.}
\label{fig:fig2}
\end{figure}

\begin{figure}[tbp]
\caption{$R_H$ vs Temperature of MgB$_2$ thin films at 5 T. Distinct
temperature dependences of the $R_H$ are evident below and above 130 K. The
data were measured by reversing the magnetic field from -5 T to 5 T at a
fixed temperature, as shown in the inset.}
\label{fig:fig3}
\end{figure}

\begin{figure}[tbp]
\caption{Temperature dependence of cot$\protect\theta _{H}$ at 5 T. A clear $%
T^{2}$ law was observed above 130 K.}
\label{fig:fig4}
\end{figure}

\begin{figure}[tbp]
\caption{Mixed-state (a) $\protect\rho _{xx}$ and (b) $\protect\rho _{xy}$
measured at applied current densities of 10$^{2}$, 10$^{3}$, and 10$^{4}$
A/cm$^{2}$ and for H = 2 and 5 T. No sign change was observed, which is
different from the case of HTS.}
\label{fig:fig5}
\end{figure}


\begin{references}
\bibitem[*]{email}  E-mail address: wnkang@postech.ac.kr

\bibitem{Nagamatsu01}  J. Nagamatsu, N. Nakagawa, T. Muranaka, Y. Zenitani,
and J. Akimitsu, Nature {\bf 410}, 63 (2001).

\bibitem{Kang01}  W. N. Kang, C. U. Jung, Kijoon H. P. Kim, Min-Seok Park,
S. Y. Lee, Hyeong-Jin Kim, Eun-Mi Choi, Kyung Hee Kim, Mun-Seog Kim, and
Sung-Ik Lee, Appl. Phys. Lett. {\bf 79}, 982 (2001).

\bibitem{Lorentz01}  B. Lorenz, R. L. Meng, Y. Y. Xue, and C. W. Chu, Phys.
Rev. {\bf B64}, 052513 (2001).

\bibitem{Choi01}  E. S. Choi, W. Kang, J. Y. Kim, Min-Seog Park, C. U. Jung,
Heon-Jung Kim, and Sung-Ik Lee, cond-mat/0104454 (2001).

\bibitem{MSKim01}  Mun-Seog Kim, C. U. Jung, Min-Seok Park, S. Y. Lee,
Kijoon H. P. Kim, W N. Kang, and Sung-Ik Lee, Phys. Rev. {\bf B64}, 012511
(2001); C. U. Jung, Min-Seok Park, W. N. Kang, Mun-Seog Kim, Kijoon H. P.
Kim, S. Y. Lee, and Sung-Ik Lee, Appl. Phys. Lett. {\bf 78}, 4157 (2001).

\bibitem{Finnemore01}  D. K. Finnemore, J. E. Ostenson, S. L. Bud'ko, G.
Lapertot, and P. C. Canfield, Phys. Rev. Lett. {\bf 86}, 2420 (2001).

\bibitem{KJKim01}  Kijoon H. P. Kim, Jae-Hyuk Choi, C. U. Jung, P.
Chowdhury, Min-Seok Park, Heon-Jung Kim, J. Y. Kim, Zhonglian Du, Eun-Mi
Choi, Mun-Seog Kim, W. N. Kang, Sung-Ik Lee, Gun Yong Sung, and Jeong Yong
Lee, cond-mat/0105330 (2001).

\bibitem{SLee01}  S. Lee, H. Mori, T. Masui, Yu. Eltsev, A. Yamamoto, and S.
Tajima, cond-mat/0105545 (2001).

\bibitem{Jung01}  M. H. Jung, M. Jaime, A. H. Lacerda, G. S. Boebinger, W.
N. Kang, Hyeong-Jin. Kim, Eun-Mi Choi, and Sung-Ik Lee, Chem. Phys. Lett. 
{\bf 343}, 447 (2001).

\bibitem{Satta01}  G. Satta, G. Profeta, F. Bernardini, A. Continenza, and
S. Massidda, Phys. Rev. {\bf B64}, 104507 (2001).

\bibitem{Jin01}  R. Jin, M. Paranthaman, H. Y. Zhai, H. M. Christen, D. K.
Christen, and D. Mandrus, cond-mat/0104411 (2001).

\bibitem{Anderson91}  P. W. Anderson, Phys. Rev. Lett. {\bf 67}, 2092 (1991).

\bibitem{Abe99}  Y. Abe, K. Segawa, and Y. Ando, Phys. Rev. {\bf B60},
R15055 (1999), and references therein.

\bibitem{Chien91}  T. R. Chien, Z. Z. Wang, and N. P. Ong, Phys. Rev. Lett. 
{\bf 67}, 2088 (1991).

\bibitem{Carrington92}  A. Carrington, A. P. Mackenzie, C. T. Lin, and J. R.
Cooper, Phys. Rev. Lett. {\bf 69}, 2855 (1992).

\bibitem{Hagen93}  S. J. Hagen, A. W. Smith, M. Rajeswari, J. L. Peng, Z. Y.
Li, R. L. Greene, S. N. Mao, X. X. Xi, S. Bhattacharya, Qi Li, and C. J.
Lobb, Phys. Rev. {\bf B47}, 1064 (1993), and references therein.

\bibitem{Kang96}  W. N. Kang, D. H. Kim, S. Y. Shim, J. H. Park, T. S. Hahn,
S. S. Choi, W. C. Lee, J. D. Hettinger, K. E. Gray, and B. Glagola, Phys.
Rev. Lett. {\bf 76}, 2993 (1996).

\bibitem{Kang00}  W. N. Kang, B. W. Kang, Q. Y. Chen, J. Z. Wu, Y. Bai, W.
K. Chu, D. K. Christen, R. Kerchner, and Sung-Ik Lee, Phys. Rev. {\bf B61},
722 (2000), and references therein.

\bibitem{Canfield01}  P. C. Canfield, D. K. Finnemore, S. L. Bud'ko, J. E.
Ostenson, G. Lapertot, C. E. Cunningham, and C. Petrovic, Phys. Rev. Lett. 
{\bf 86}, 2423 (2001).

\bibitem{Kang01b}  W. N. Kang, Hyeong-Jin Kim, Eun-Mi Choi, C. U. Jung, and
Sung-Ik Lee, Science {\bf 252}, 1521 (2001); (10.1126/science 1060822).

\bibitem{HJKim01}  Hyeong-Jin Kim, W. N. Kang, Eun-Mi Choi, Mun-Seog Kim,
Kijoon H. P. Kim, and Sung-Ik Lee, Phys. Rev. Lett. {\bf 87}, 087002 (2001).

\bibitem{Harris92}  J. M. Harris, Y. F. Yan, and N. P. Ong, Phys. Rev. {\bf %
B46}, 14293 (1992).

\bibitem{HJKim01b}  Heon-Jung Kim, W. N. Kang, Hyeong-Jin Kim, Eun-Mi Choi,
Kijoon H. P. Kim, H. S. Lee, and Sung-Ik Lee (unpublished).

\bibitem{Manzano01}  F. Manzano, A. Carrington, N. E. Hussey, S. Lee, and A.
Yamamoto, cond-mat/0110109 (2001).

\bibitem{Dorsey92}  A. T. Dorsey, Phys. Rev. B ${\bf 46}$, 8376 (1992).

\bibitem{Kopnin93}  N. B. Kopnin, B. I. Ivlev, and V. A. Kalatsky, J. Low
Temp. Phys. ${\bf 90}$, 1 (1993); N. B. Kopnin and A. V. Lopatin, Phys. Rev.
B ${\bf 51}$, 15291 (1995); N. B. Kopnin, Phys. Rev. B ${\bf 54}$, 9475
(1996)..

\bibitem{Otterlo95}  A. van Otterlo, M. Feigel'man, V. Geshkenbein, and G.
Blatter , Phys. Rev. Lett. ${\bf 75}$, 3736 (1995).

\bibitem{Ginsberg95}  D. M. Ginsberg and J. T. Manson, Phys. Rev. B ${\bf 51}
$, 515 (1995); C. C. Almasan, S. H. Han, K. Yoshiara, M. Buchgeister, D. A.
Gajewski, L. M. Paulius, J. Herrmann, M. B. Maple, A. P. Paulikas, Chun Gu,
and B. W. Veal, {\it ibid}. ${\bf 51}$, 3981 (1995); J. T. Kim, J.
Giapintzakis, and D. M. Ginsberg, {\it ibid}. ${\bf 53}$, 5922 (1996).
\end{references}
\end{document}